\documentclass[10pt]{article}
\usepackage{graphicx}

\begin{document}
\title{Gravitational Waves and the Sagnac Effect}
\author{S. Sivasubramanian, A. Widom, Y.N. Srivastava \\
Physics Department, Northeastern University, Boston MA USA \\
and \\
Physics Department \& INFN, University of Perugia, Perugia Italy}
\date{}
\maketitle
\begin{abstract}
We consider light waves propagating clockwise and other light waves
propagating counterclockwise around a closed path in a plane
(theoretically with the help of stationary mirrors). The time difference
between the two light propagating path orientations constitutes the Sagnac effect.
The general relativistic expression for the Sagnac effect is discussed. It is
shown that a gravitational wave incident to the light beams at an
arbitrary angle will {\em not} induce a Sagnac effect so long as the wave length of
the weak gravitational wave is long on the length scale of the closed light beam
paths. The gravitational wave induced Sagnac effect is thereby null.
\end{abstract}

\section{Introduction \label{intro}}
With the help of reflecting mirrors, one may propagate a light signal in a clockwise
closed path \begin{math} (C) \end{math}. Similarly, a light signal can be propagated
in the counterclockwise closed path \begin{math} (-C) \end{math}.
The time of propagation in the clockwise orientation need not be equal to the time of
propagation in the counterclockwise direction. The time difference
\begin{math} \Delta t =t[C]-t[-C]\end{math} between the two possible oriented
paths constitutes the Sagnac effect. The original experimental
verification\cite{Sagnac:1913} of this effect required that the mirrors be
fixed to a rotating table. Two light beams going in opposite directions exhibited a
phase interference depending linearly on the angular velocity of the table.
If the path \begin{math} C \end{math} is the boundary of a plane open surface
\begin{math} \Sigma \end{math}, i.e. \begin{math} C=\partial \Sigma \end{math}, then
the time difference is given by
\begin{equation}
\Delta t=\frac{4}{c^2}\int_\Sigma {\bf \Omega }\cdot d{\bf \Sigma},
\label{intro1}
\end{equation}
where \begin{math} {\bf \Omega } \end{math} is the axial vector angular velocity.
The Sagnac effect forms the basis of practical engineering devices measuring
rotational velocities.

\begin{figure}[tp]
\centering
\includegraphics[width=3.5in]{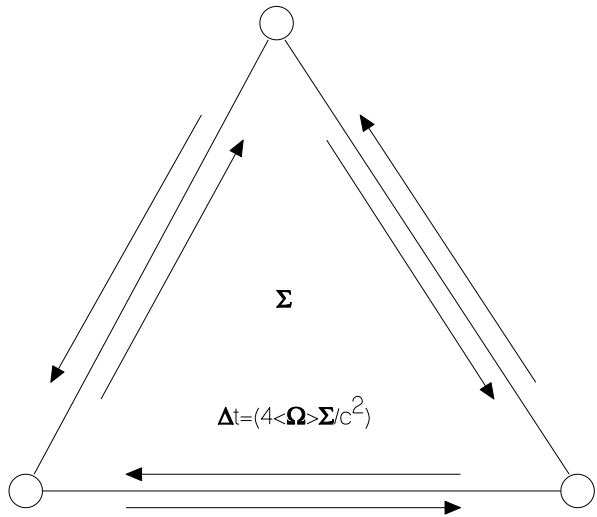}
\caption{In a triangular geometrical arrangement contemplated for LISA,
one may consider the mean flux of the local axial angular velocity vector
${\bf \Omega}$ through the area $\Sigma $. In detail,
$\int_\Sigma {\bf \Omega }\cdot d{\bf \Sigma }=\left<\Omega \right>\Sigma $
yields the time difference between the clockwise and counterclockwise light
ray paths according to the Sagnac rule $\Delta t=(4\left<\Omega \right>\Sigma/c^2)$.
A plane polarized gravitational wave does indeed produce a local rotational
axial vector ${\bf \Omega }$. However, in the limit in which the wavelength
$\lambda $ of the gravitational wave obeys $\lambda^2>>\Sigma $ and to lowest
order in the gravitational wave curvature, one finds that
$\left<\Omega \right>\approx 0$.
The Sagnac effect is thereby {\em null}.}
\label{fig1}
\end{figure}

More recently, proposals that the Sagnac effect may be made the basis of
gravitational wave
detection\cite{Tinto:1995,Hellings:1996,Folkner:1997,Bender:1998,Armstrong:1999}
have been of considerable recent
interest\cite{Tinto:2000,Estabrook:2000,Hellings:2001,Hogan:2001}.
The experimental technique envisions a
{\em Laser Interferometer Space Antenna} (LISA). The general
idea\cite{Daniel:2003,Estrabrook:2003,Tinto:2003} is illustrated
in {\em schematic} form in Fig.\ref{fig1}. Three
enclosed reflecting mirrors float in space forming a triangle. Clockwise and
counterclockwise beams of light propagate on a path \begin{math} C \end{math}
which bounds a triangular area \begin{math} \Sigma \end{math}. Let
\begin{math} L \end{math} denote the length scale of the sides of the
triangle and let \begin{math} \lambda \end{math} denote the wavelength
of a polarized gravitational wave incident upon the triangle from a
direction described by a unit vector \begin{math} {\bf n} \end{math}. It
is proposed that the Sagnac time difference in Eq.(\ref{intro1}) in the limit
\begin{math} \lambda >> L \end{math} can be employed to probe the local
rotational velocity \begin{math} {\bf \Omega} \end{math} produced by the
gravitational wave. The principle is that one may measure the flux of
the rotational velocity
\begin{math} \int_\Sigma {\bf \Omega}\cdot d{\bf \Sigma}  \end{math}.

In order to see what is involved, let us consider an
electromagnetic and gravitational wave analogy.
Suppose a measuring device for magnetic flux
\begin{math}\Phi =\int_\Sigma {\bf B}\cdot d{\bf \Sigma}  \end{math}
and further suppose that one wished to detect
a plane polarized electromagnetic wave by measuring the magnetic field
\begin{math} {\bf B} \end{math} via a magnetic flux reading device.
If the electromagnetic wavelength and the flux area obey
\begin{equation}
\lambda^2 >> \Sigma ,
\label{intro2}
\end{equation}
then the magnetic flux
\begin{math}\Phi =\int_\Sigma {\bf B}\cdot d{\bf \Sigma}  \end{math}
meter could serve as a polarized electromagnetic wave detector provided that
the direction \begin{math} {\bf n} \end{math} and polarization
\begin{math} {\bf e} \end{math} of the electromagnetic wave were in appropriate
directions relative to the flux area \begin{math} \Sigma \end{math}.
Similarly, for a gravitational wave incident upon the
LISA arrangement pictured in Fig.\ref{fig1}, the Sagnac time difference
\begin{math}\Delta t =(4/c^2)\int_\Sigma {\bf \Omega }\cdot d{\bf \Sigma}  \end{math}
measures the flux of a gravitational wave induced rotational velocity
\begin{math} {\bf \Omega } \end{math} through the area
\begin{math} \Sigma \end{math}. However, it will
be shown in what follows, that if the gravitational wavelength
\begin{math} \lambda \end{math} obeys Eq.(\ref{intro2}), then to linear order
in the gravitational wave curvature tensor the Sagnac
\begin{math} \Delta t \approx 0 \end{math}. The LISA gravitational wave induced
Sagnac effect is thereby theoretically null.

In Sec.\ref{se}, the general expression for the Sagnac effect in a quasi-static
metric is derived. A (possibly strong) gravitational plane wave solution to
the vacuum Einstein equations is explored in Sec.\ref{gw}. An expression for
the local rotational velocity \begin{math} {\bf \Omega} \end{math} due
to a gravitational wave is derived. In Sec.\ref{rvf}, the Sagnac delay time
\begin{math} \Delta t  \end{math} due to a quasi-static gravitational wave
is computed and shown to be null in the quasi-static limit of Eq.(\ref{intro2}).
In the concluding Sec.{\ref{con}}, the symmetry principles involved in
the theory are discussed from a physical viewpoint.

\section{Sagnac Effect \label{se}}
Consider a general proper time metric in general relativity
\begin{equation}
-c^2 d\tau^2=g_{\mu \nu }dx^\mu dx^\nu
\label{se1}
\end{equation}
wherein
\begin{math}
x=(x^1,x^2,x^3,x^0)=(r^1,r^2,r^3,ct)=({\bf r},ct)
\end{math}
denotes space time coordinates. Employing the definitions
\begin{eqnarray}
g_{00}(x) &=& -h({\bf r},t),
\nonumber \\
g_{i0}(x) &=& g_{0i}(x)=\frac{h({\bf r},t)u_i({\bf r},t)}{c}\ ,
\nonumber \\
g_{ij}(x) &=& \gamma_{ij}({\bf r},t)-h({\bf r},t)
\left(\frac{u_i({\bf r},t)u_j({\bf r},t)}{c^2}\right),
\label{se2}
\end{eqnarray}
yields the metric expressions
\begin{eqnarray}
c^2d\tau ^2 &=& c^2 d\tilde{t}^{\ 2}-d\ell ^2,
\nonumber \\
c d\tilde{t} &=&\sqrt{h}\left(c dt-\frac{u_idr^i}{c}\right)
=\sqrt{h}\big(c dt-({\bf u\cdot }d{\bf r}/c)\big)
\nonumber \\
d\ell ^2 &=& \gamma_{ij}dr^i dr^j={\bf \gamma }:d{\bf r}d{\bf r} .
\label{se3}
\end{eqnarray}
The metric \begin{math} \{\gamma_{ij}\} \end{math} will be used
in what follows to describe the possibly curved spatial geometry
via the distance \begin{math} d\ell \end{math}.

A light signal connecting two neighboring events is characterized by
a zero proper time interval or equivalently by a propagation
light speed \begin{math} c \end{math}; i.e.
\begin{equation}
d\tau =0\ \ \ {\rm implies}
\ \ \ \left|\frac{d\ell }{d\tilde{t}}\right|=c.
\label{se4}
\end{equation}
The following is worthy of note: If one reverses the spatial displacement
between two neighboring events
\begin{math} d{\bf r}\to -d{\bf r} \end{math}
both connected by light signals, then the change
in the effective time of those signals obeys
\begin{math} \Delta(dt)=2({\bf u}\cdot d{\bf r}/c^2)\end{math}.
In more detail, let us suppose that light (with the help of perfectly
reflecting mirrors) moved around a closed path \begin{math} C \end{math}.
Let us also suppose that different light moved ``backwards'' around the same
closed path. If the metric is {\em quasi-static} on the time scale
of the round trip \begin{math} C \end{math} and the reversed round trip
\begin{math} -C \end{math}, then the time difference between the two trips
obeys
\begin{equation}
\Delta t=\frac{1}{c^2}\left(\oint_C -\oint_{-C}\right) u_idr^i=
\frac{2}{c^2}\oint_C u_idr^i\equiv
\frac{2}{c^2}\oint_{C=\partial \Sigma}{\bf u}\cdot d{\bf r},
\label{se5}
\end{equation}
where we have noted that the closed curve
\begin{math} C=\partial \Sigma \end{math}
is the boundary of an open surface \begin{math} \Sigma \end{math}.
Using the spatial metric \begin{math} d\ell ^2=\gamma_{ij}dr^i dr^j \end{math}
(\begin{math} \gamma=\det{\gamma_{ij}} \end{math}), allows us to define
\begin{math} {\bf \Omega } \end{math} via
\begin{equation}
\partial_i u_j-\partial_j u_i =\sqrt{\gamma }\ \epsilon_{ijk}
({\bf curl\ u})^k ,
\ \ \ \ d{\bf \Sigma}_k =
\frac{1}{2}\sqrt{\gamma}\ (dr^i \wedge dr^j)\epsilon_{ijk},
\label{se6}
\end{equation}
\begin{equation}
{\bf \Omega}=\frac{1}{2}{\bf curl\ u}.
\label{se7}
\end{equation}
\medskip
\par \noindent
{\bf Theorem 1:\ (The Sagnac Effect)}
\medskip
\par \noindent
{\em Consider a light beam moving clockwise and another light beam moving
counterclockwise around a closed curve boundary $C=\partial \Sigma$ of an
open surface $\Sigma $ in a quasi-static gravitational metric.
The difference in the circulation times is proportional to the flux of the
local angular velocity ${\bf \Omega}$ through $\Sigma $};
\begin{equation}
\Delta t=\frac{4}{c^2}\int_\Sigma \Omega^k d\Sigma_k
=\frac{4}{c^2}\int_\Sigma {\bf \Omega }\cdot d{\bf \Sigma }.
\label{se8}
\end{equation}
{\bf Proof:} {\em Eq.(\ref{se8}) follows from Eqs.(\ref{se5}) and (\ref{se7})
using Stokes theorem.}
\medskip

Previous experimental verifications of this theorem have included measurements
of the angular velocity of a spinning apparatus and the angular velocity of the
earth. The experimental verifications required phase interference between
the clockwise and counterclockwise paths for electromagnetic waves.
The Sagnac measured interference phase shift
\begin{math} \Delta \Theta =\omega \Delta t \end{math}
for light of frequency \begin{math} \omega \end{math}
is given by
\begin{math}
\Delta \Theta =(4\omega /c^2)\int_\Sigma {\bf \Omega }\cdot d{\bf \Sigma }.
\end{math}
In the application of the Sagnac effect to deep space probes such
as LISA, the time difference \begin{math} \Delta t  \end{math}
{\em induced by a gravitational wave} is the central issue.

\section{Gravitational Waves \label{gw}}
A gravitational plane wave, which is not necessarily weak, may be theoretically
described as follows\cite{Stephani:2002}:  (i) Start from flat space time with
the metric
\begin{equation}
-c^2d\tau_0^2=-c^2dt^2+|d{\bf r}|^2.
\label{gw1}
\end{equation}
(ii) Choose a unit spatial vector \begin{math} {\bf n} \end{math} for the propagation
direction of the gravitational wave and define spatial coordinates parallel
\begin{math} r_\| \end{math} and normal \begin{math} {\bf r}_\perp \end{math}
to the propagation direction; i.e.
\begin{equation}
{\bf n \cdot n}=1, \ \ \ r_{\|}={\bf n \cdot r}
\ \ \ {\rm and}\ \ \ {\bf r}_\perp={\bf n}\times ({\bf r}\times {\bf n}).
\label{gw2}
\end{equation}
(iii) The gravitational wave disturbance (at first described as a dimensionless ``scalar''
\begin{math} W \end{math}) obeys the wave equation
\begin{equation}
\left\{\frac{1}{c^2}\left(\frac{\partial}{\partial t}\right)^2
-\Delta \right\}W({\bf r},t)=
\left\{\frac{1}{c^2}\left(\frac{\partial}{\partial t}\right)^2
-\left(\frac{\partial}{\partial r_\|}\right)^2
-\Delta_\perp \right\}W({\bf r},t)=0,
\label{gw3}
\end{equation}
where \begin{math} \Delta W\equiv (div\ {\bf grad})W \end{math}. (iv) The {\em plane
wave} solution of Eq.(\ref{gw3}) has the form
\begin{equation}
W({\bf r},t) = \Psi \big({\bf r}_\perp ,t-(r_\|/c)\big)
=\Psi \big({\bf r}_\perp ,t-({\bf n\cdot r}/c)\big),
\label{gw4}
\end{equation}
with \begin{math} \Psi \end{math} obeying Laplace's equation in the
normal plane coordinates
\begin{equation}
\Delta_\perp \Psi ({\bf r}_\perp ,t) = 0.
\label{gw5}
\end{equation}
(v) Finally, the proper time in the presence of the gravitational wave
is given by
\begin{eqnarray}
-c^2d\tau^2 &=& -c^2d\tau_0^2-W(dr_\|-c dt)^2,
\nonumber \\
-c^2d\tau^2 &=&-c^2 dt^2+|d{\bf r}_\perp|^2+dr_\|^2-W(dr_\|-c dt)^2,
\nonumber \\
-c^2d\tau^2 &=&-(1+W)c^2 dt^2+|d{\bf r}_\perp|^2+(1-W)dr_\|^2
+2Wcdtdr_\|.
\label{gw6}
\end{eqnarray}
The resulting proper time metric
\begin{equation}
-c^2d\tau^2 =-(1+W)\left\{cdt-\left(\frac{W}{1+W}\right)dr_\|\right\}^2
+|d{\bf r}_\perp|^2+\left\{\frac{dr_\|^2}{1+W}\right\},
\label{gw7}
\end{equation}
with \begin{math} W \end{math} given in Eqs.(\ref{gw4}) and (\ref{gw5}),
represents an {\em exact gravitational wave solution} to the vacuum Einstein
curvature equation
\begin{math} {\cal R}_{\mu \nu}={\cal R}^{\lambda }_{\ \ \mu \lambda \nu}=0 \end{math}.
There is no {\em theoretical} requirement that the gravitational disturbance be small
in the sense that \begin{math} |W|\ll 1 \end{math}. However it is expected that
{\em experimental} gravitational wave disturbances will be very small indeed.

From Eqs.(\ref{se3}), (\ref{gw2}) and (\ref{gw7}) it follows that a gravitational
wave carries a local flow velocity
\begin{equation}
{\bf u}=\frac{cW{\bf n}}{1+W}\ .
\label{gw8}
\end{equation}
and local rotational velocity
\begin{eqnarray}
{\bf \Omega }({\bf r},t) &=& \frac{\bf curl\ u({\bf r},t)}{2}\ ,
\nonumber \\
{\bf \Omega }({\bf r},t) &=& \frac{c}{2}
\left\{\frac{{\bf grad}W({\bf r},t){\bf \times n}}{(1+W({\bf r},t))^2}\right\},
\nonumber \\
{\bf \Omega }({\bf r},t) &=& \frac{c}{2}
\left\{\frac{{\bf grad}_\perp \Psi \big({\bf r}_\perp ,t-(r_\|/c)\big){\bf \times n}}
{\left[1+\Psi \big({\bf r}_\perp ,t-(r_\|/c)\big)\right]^2}\right\}.
\label{gw9}
\end{eqnarray}
With the spatial length scale
\begin{eqnarray}
d\ell ^2 &=& \gamma_{ij}dr^idr^j=
|d{\bf r}_\perp |^2+\left\{\frac{dr_\|^2}{1+W}\right\}\ ,
\nonumber \\
\gamma  &=& \det(\gamma_{ij})=\frac{1}{1+W}\ ,
\label{gw10}
\end{eqnarray}
the cross product in Eqs.(\ref{gw9}) associated with the spatial metric
is given by
\begin{equation}
({\bf a\times b})^k=\left(\frac{a_i b_j}{\sqrt{\gamma }}\right)\epsilon^{ijk} .
\label{gw11}
\end{equation}

The Sagnac effect time delay \begin{math} \Delta t \end{math} induced by a
gravitational wave is rigorously determined by Eqs.(\ref{se8}), (\ref{gw5})
and (\ref{gw9}). Let us examine the consequences of the above results for
a plane polarized gravitational wave. The problem at hand is to find
an appropriate solution to Eq.(\ref{gw5}).

\section{Gravitational Wave Rotational Velocity Flux \label{rvf}}

Let us assume that before the gravitational wave impinges on the
laser interferometer space antenna. The center of area
\begin{math} {\bf R} \end{math} of the triangle is given by
\begin{equation}
{\bf R}\ \Sigma = \int_{\Sigma }{\bf r}\ d\Sigma .
\label{rvf1}
\end{equation}
If \begin{math} {\sf e}  \end{math} denotes the symmetric polarization
tensor of the gravitational wave, then we have
\begin{equation}
{\sf e}\cdot {\bf n} = {\bf n}\cdot {\sf e}=0\ \ {\rm and}
\ \ tr\{{\sf e}\}=0.
\label{rvf2}
\end{equation}
The solution to Eq.(\ref{gw5}) corresponding to gravitational wave
reads
\begin{equation}
\Psi ({\bf r}_\perp ,t) = \{{\bf r}-{\bf R}\}_\perp
\cdot {\sf e} \cdot \{{\bf r}-{\bf R}\}_\perp \ {\cal A}(t),
\label{rvf3}
\end{equation}
where the non-vanishing curvature tensor elements are proportional
to the amplitude \begin{math} {\cal A}\big(t-(r_\|/c)\big)  \end{math}.
Eqs.(\ref{gw9}) and (\ref{rvf3}) imply (for a weak gravitational wave)
\begin{equation}
{\bf \Omega } = c\ {\sf e} \cdot \{{\bf r}-{\bf R}\}_\perp {\bf \times}
{\bf n}\ {\cal A}\big(t-(r_\|/c)\big)+\ldots \ .
\label{rvf4}
\end{equation}
Thus, to lowest order in the gravitational wave curvature and to lowest
order in the gravitational wave frequency
\begin{math} (c/\lambda ) \end{math} we find that
\begin{eqnarray}
c\Delta t &=& \frac{4}{c}\int_\Sigma {\bf \Omega }\cdot d{\bf \Sigma},
\nonumber \\
c\Delta t & \approx & 4{\cal A}(t) \int_\Sigma {\sf e} \cdot
\{{\bf r}-{\bf R}\}_\perp {\bf \times } {\bf n}\cdot d{\bf \Sigma },
\nonumber \\
c\Delta t & \approx & 4{\cal A}(t)\ {\sf e} \cdot\left(
\int_\Sigma ({\bf r-R})d\Sigma
\right) {\bf \times }{\bf n}\cdot {\bf N},
\label{rvf5}
\end{eqnarray}
wherein the unit vector \begin{math} {\bf N} \end{math} is normal to
the area \begin{math} d{\bf \Sigma} = {\bf N} d\Sigma \end{math}.
\medskip
\par \noindent
{\bf Theorem 2:\ (The Gravitational Wave Sagnac Effect)}
\medskip
\par \noindent
{\em The quasi-static $(\lambda^2 >> \Sigma )$ weak $(|W|<<1)$
gravitational wave Sagnac effect is null};
\begin{equation}
\Delta t \approx 0.
\label{rvf6}
\end{equation}
{\bf Proof:} {\em Eq.(\ref{rvf6}) follows from Eqs.(\ref{rvf1}) and (\ref{rvf5}).}
\medskip
\par \noindent
The above theorem is the central result of this work.

\section{Conclusions \label{con}}

Let us consider the physical meaning of the local drift velocity
\begin{math} {\bf u}  \end{math}. If we examine Eq.(\ref{se3}), then
the null proper time \begin{math} d\tau =0  \end{math} of a massless photon
traveling on a light ray path yields a light ray optical path length
\begin{equation}
dL = cdt = \frac{d\ell }{\sqrt{h}}+\frac{\bf u}{c}\cdot d{\bf r}.
\label{con1}
\end{equation}
For a path \begin{math} {\cal P} \end{math} in a quasi-static spatial metric
\begin{math} d\ell ^2=\gamma_{ij}dr^i dr^j \end{math} one may compute the optical
path length
\begin{equation}
L[{\cal P}] =c\int_{\cal P} dt = \int_{\cal P}
\left(\frac{d\ell }{\sqrt{h}}+\frac{\bf u}{c}\cdot d{\bf r}\right).
\label{con2}
\end{equation}
The Fermat principle of least time,
\begin{math} \delta \int_{\cal P}dt =0 \end{math}
yields the principle of least optical path length
\begin{equation}
\delta L[{\cal P}] = 0.
\label{con3}
\end{equation}
Eqs.(\ref{con2}) and (\ref{con3}) yield the light ray equation in a quasi-static
gravitational field. That the Fermat variational principle Eq.(\ref{con3}) yields
experimental light propagation is evident ever since the very first experiments on
the gravitational bending of light rays. In Eq.(\ref{con2}), the index of refraction
of the gravitational field
\begin{equation}
n=\frac{1}{\sqrt{h}}\ , \ \ {\rm yields}
\ \  L_0[{\cal P}]=\int_{\cal P}\frac{d\ell }{\sqrt{h}}=\int_{\cal P}nd\ell .
\label{con4}
\end{equation}
The additional optical path length (\begin{math} L=L_0+\Delta L \end{math}),
as given by
\begin{equation}
\Delta L[{\cal P}] =c\Delta t[{\cal P}] = \int_{\cal P}
\frac{\bf u}{c}\cdot d{\bf r},
\label{con5}
\end{equation}
is due to the local drift velocity \begin{math} {\bf u} \end{math}.
The meaning of the local drift velocity becomes evident if we consider two
photons with equal but opposite momenta
\begin{equation}
{\bf p}_+ = \hbar {\bf k}\ \ \ {\rm and}
\ \ \ {\bf p}_- = -\hbar {\bf k}.
\label{con6}
\end{equation}
The velocities of the two photons, \begin{math} {\bf v}_+ \end{math}
and \begin{math} {\bf v}_- \end{math}, will not {\em quite} be equal and opposite.
In the notation of Eq.(\ref{se2}), we have
\begin{equation}
({\bf v}_+ +{\bf v}_-) =
\left\{\frac{2h{\bf u}}{1-h|({\bf u}/c)|^2}\right\}
\approx 2{\bf u} \approx 2cW{\bf n}
\label{con7}
\end{equation}
to linear order in the gravitational wave amplitude
\begin{math} W \end{math}.

A photon moving parallel to the gravitational wave need not have the
same speed as a photon moving anti-parallel to the gravitational wave
since the drift velocity
\begin{math} {\bf u}=\{cW{\bf n}/(1+W)\}  \end{math}
is non-vanishing is some regions of space. The photon velocity
in the clockwise and counterclockwise orientations will have along the
two paths (in general) \begin{math} {\bf v}_+\ne -{\bf v}_-  \end{math}.
However, for a plane gravitational wave incident upon the triangle
with \begin{math} W({\bf R},t)=0\end{math}, we have varying directions for
\begin{math} ({\bf v}_+ + {\bf v}_-)  \end{math}. The resulting flux
\begin{math}
(c^2\Delta t/4)=\int_\Sigma {\bf \Omega }\cdot d{\bf \Sigma }\approx 0.
\end{math}
Experiments which measure gravitational waves via the Sagnac effect
are thereby rendered unlikely to succeed.

\end{document}